\documentclass[twocolumn,superscriptaddress,showpacs,preprintnumbers,amsmath,amssymb,pre]{revtex4}
\usepackage{graphicx}
\usepackage{dcolumn,xcolor,ulem}
\usepackage{amsmath} 
\usepackage{amssymb}
\usepackage{amsfonts}
\usepackage{bm}
\usepackage[latin1]{inputenc}
\usepackage{hyperref} 

\begin{document}


 \title{Wall slip across the jamming transition of soft thermoresponsive particles}

\author{Thibaut Divoux}
\email{divoux@crpp-bordeaux.cnrs.fr}
\affiliation{Centre de Recherche Paul Pascal, CNRS UPR 8641 - 115 avenue Dr. Schweitzer, 33600 Pessac, France}
\author{V\'eronique Lapeyre}
\affiliation{Universit\'e de Bordeaux, ISM, ENSCBP, 33607 Pessac Cedex, France}
\author{Val\'erie Ravaine}
\affiliation{Universit\'e de Bordeaux, ISM, ENSCBP, 33607 Pessac Cedex, France}
\author{S\'ebastien Manneville}
\affiliation{Universit\'e de Lyon, Laboratoire de Physique, \'Ecole Normale Sup\'erieure de Lyon, CNRS UMR 5672, 46 All\'ee d'Italie, 69364 Lyon cedex 07, France}

\date{\today}

\begin{abstract}
Flows of suspensions are often affected by wall slip, that is the fluid velocity $v_{f}$ in the vicinity of a boundary differs from the wall velocity $v_{w}$ due to the presence of a lubrication layer. While the slip velocity $v_s=\vert v_{f}-v_{w}\vert$ robustly scales linearly with the stress $\sigma$ at the wall in dilute suspensions, there is no consensus regarding denser suspensions that are sheared in the bulk, for which slip velocities have been reported to scale as a $v_s\propto\sigma^p$ with exponents $p$ inconsistently ranging between 0 and 2. Here we focus on a suspension of soft thermoresponsive particles and show that $v_s$ actually scales as a power law of the viscous stress $\sigma-\sigma_c$, where $\sigma_c$ denotes the yield stress of the bulk material. By tuning the temperature across the jamming transition, we further demonstrate that this scaling holds true over a large range of packing fractions $\phi$ on both sides of the jamming point and that the exponent $p$ increases continuously with $\phi$, from $p=1$ in the case of dilute suspensions to $p=2$ for jammed assemblies. These results allow us to successfully revisit inconsistent data from the literature and paves the way for a continuous description of wall slip above and below jamming.
\end{abstract}

\maketitle

\section{Introduction}

Wall slip is ubiquitous in flows of complex fluids and plays a key role in numerous situations ranging from turbulent drag reduction to extrusion processes \cite{Denn:2001}. It denotes a situation where the fluid velocity $v_f$ in the vicinity of a boundary differs from the velocity $v_w$ of the boundary itself, and often originates from a chemical or physical mismatch between the constituents of the fluid and that of the wall. Such apparent discontinuity in velocity at a boundary reflects the existence of a thin and highly sheared region adjacent to the wall where the viscosity is much smaller than in the bulk material \cite{Barnes:1995}. 
Although historically described as a mere artifact, wall slip appears as an intrinsic feature of complex fluids that has been reported in polymeric fluids \cite{Lettinga:2009,Feindel:2010,Fardin:2012} as well as in soft glassy materials such as colloidal suspensions \cite{Persello:1994,Ballesta:2012}, foams \cite{Denkov:2005,Marze:2008,Cantat:2013}, emulsions \cite{Salmon:2003,Seth:2012,Mansard:2014}, microgels \cite{Magnin:1990,Meeker:2004a}, attractive gels \cite{Grenard:2014}, etc. Moreover, wall slip is often non-trivially coupled to the bulk dynamics of complex fluids, especially during the yielding transition where it may even govern the nature of the subsequent steady-state flow \cite{Gibaud:2008,Martin:2012,Kurokawa:2015}.

Physical insights into wall slip are generally gained from the scaling of the slip velocity $v_s=\vert v_{f}-v_{w}\vert$ with the stress at the wall $\sigma$. More specifically for particulate suspensions below jamming, the particle deformability does not play a key role and slip velocities scale roughly linearly with $\sigma$ \cite{Yilmazer:1989,Yilmazer:1991,Salmon:2003,Davies:2008}. In jammed particulate assemblies, however, the situation is much more subtle as the material exhibits a yield stress $\sigma_c$ below which it cannot flow in the bulk. The main goal of the present Rapid Communication is to clarify the slip behavior of a jammed suspension of soft particles driven above the yield stress. Indeed in this case $v_s$ has been reported to increase as a quadratic function of $\sigma$ in concentrated emulsions and microgels \cite{Salmon:2003,Geraud:2013} while other experiments on similar samples rather led to a linear scaling of $v_s$ with $\sigma$ \cite{Seth:2012,PerezGonzalez:2012,Poumaere:2014}. 
 
Therefore despite a large body of work on wall slip, the key parameters controlling the scaling of $v_s$ in a dense, yielded suspension are still unknown. One of these parameters is the packing fraction which effect on slip has been scarcely investigated. Here we focus on a suspension of thermoresponsive poly($N$-isopropylacrylamide) particles that allows us to explore the slip behavior of a suspension at various packing fractions in a single set of experiments, from a dilute system up to a jammed assembly of soft particles. Extensive rheological experiments coupled to velocimetry demonstrate that the slip velocity scales as a power-law on both sides of the jamming transition. For low packing fractions we recover $v_s \propto \sigma^p$ with $p\simeq 1$, while above jamming and for stresses larger than the yield stress, $v_s$ follows a power law of the viscous stress, i.e. $v_s \propto (\sigma-\sigma_c)^p$. Furthermore the exponent $p$ increases continuously from $p=1$ to $p=2$ as the packing fraction is increased. This is confirmed by revisiting formerly published data on emulsions and microgels. Our results thus explain long-standing discrepancies in the literature and bridge the gap between the slip behavior of dilute suspensions and soft jammed materials.

\begin{figure}[!t]
\centering
\includegraphics[width=\linewidth]{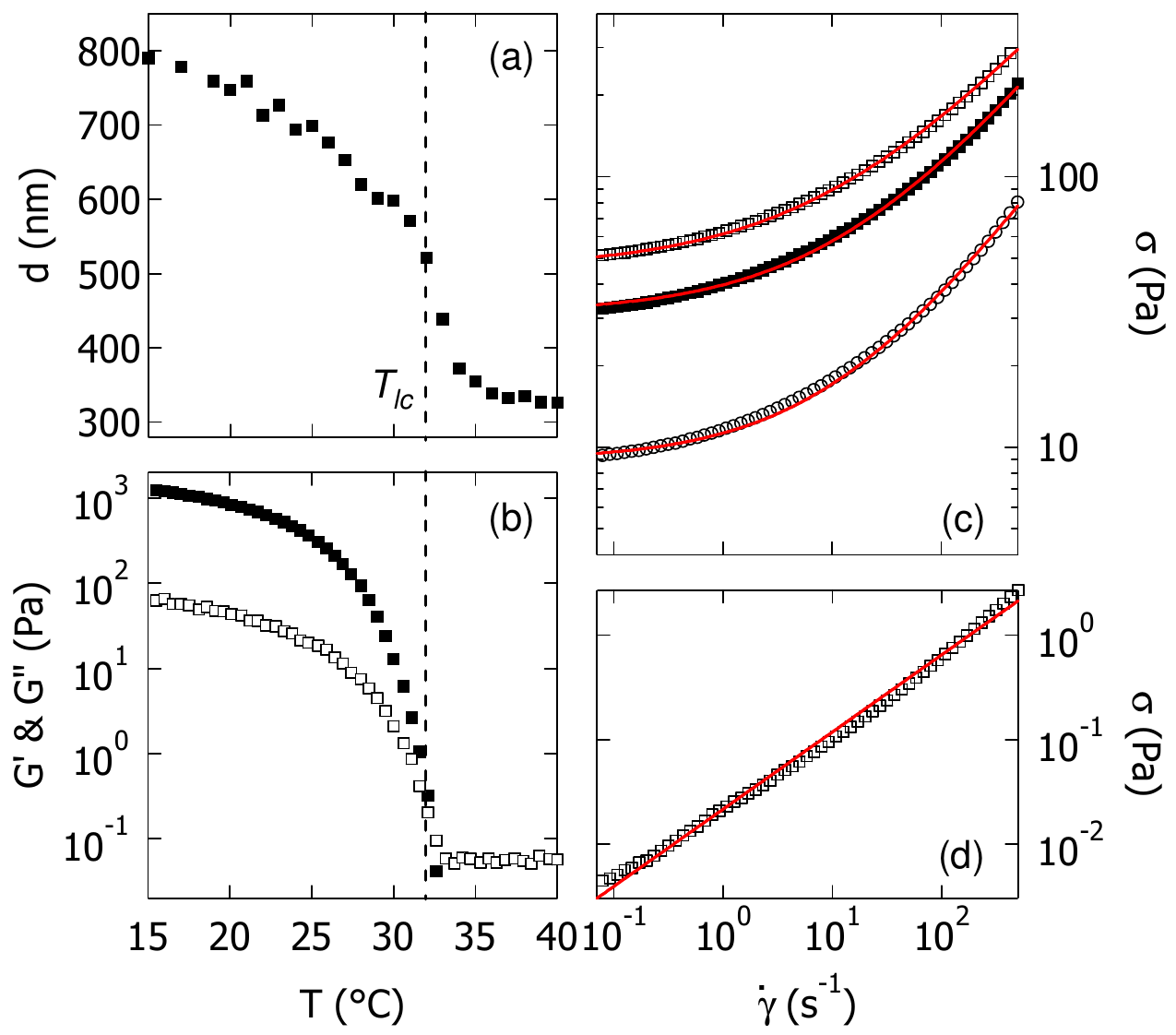}
\caption{(color online) (a)~Average diameter of the pNIPAm particles vs temperature $T$ as measured by dynamic light scattering. (b)~Elastic ($G'$, $\blacksquare$) and viscous ($G''$, $\square$) moduli measured by small amplitude oscillatory shear ($\gamma=1$~\% and $\omega=1$~Hz). The data correspond to the average of an increasing and decreasing ramp of temperature of duration 25~min each. (c) and (d)~Flow curves, shear stress $\sigma$ versus shear rate $\dot \gamma$, recorded at different temperatures: $T=17$, 20, 26 and 34$^{\circ}$C from top to bottom. Each flow curve is obtained by sweeping down $\dot \gamma$ from 500 to $5.10^{-3}$~s$^{-1}$ with a waiting time  of 10~s per point. The red curves are the best fits of the data by the Herschel-Bulkley model with $\sigma_c=45.7 \pm 0.2$, $31.3\pm 0.4$ and $8.9\pm0.1$~Pa from top to bottom in (c) and by a power law fit, $\sigma= \tilde{\eta}\dot \gamma^{n}$, with $n=0.74$ and $\tilde{\eta}=0.0245$~Pa.s$^{0.74}$ in (d). 
\label{fig1}}
\end{figure} 

\begin{figure}[b]
\centering
\includegraphics[width=0.9\linewidth]{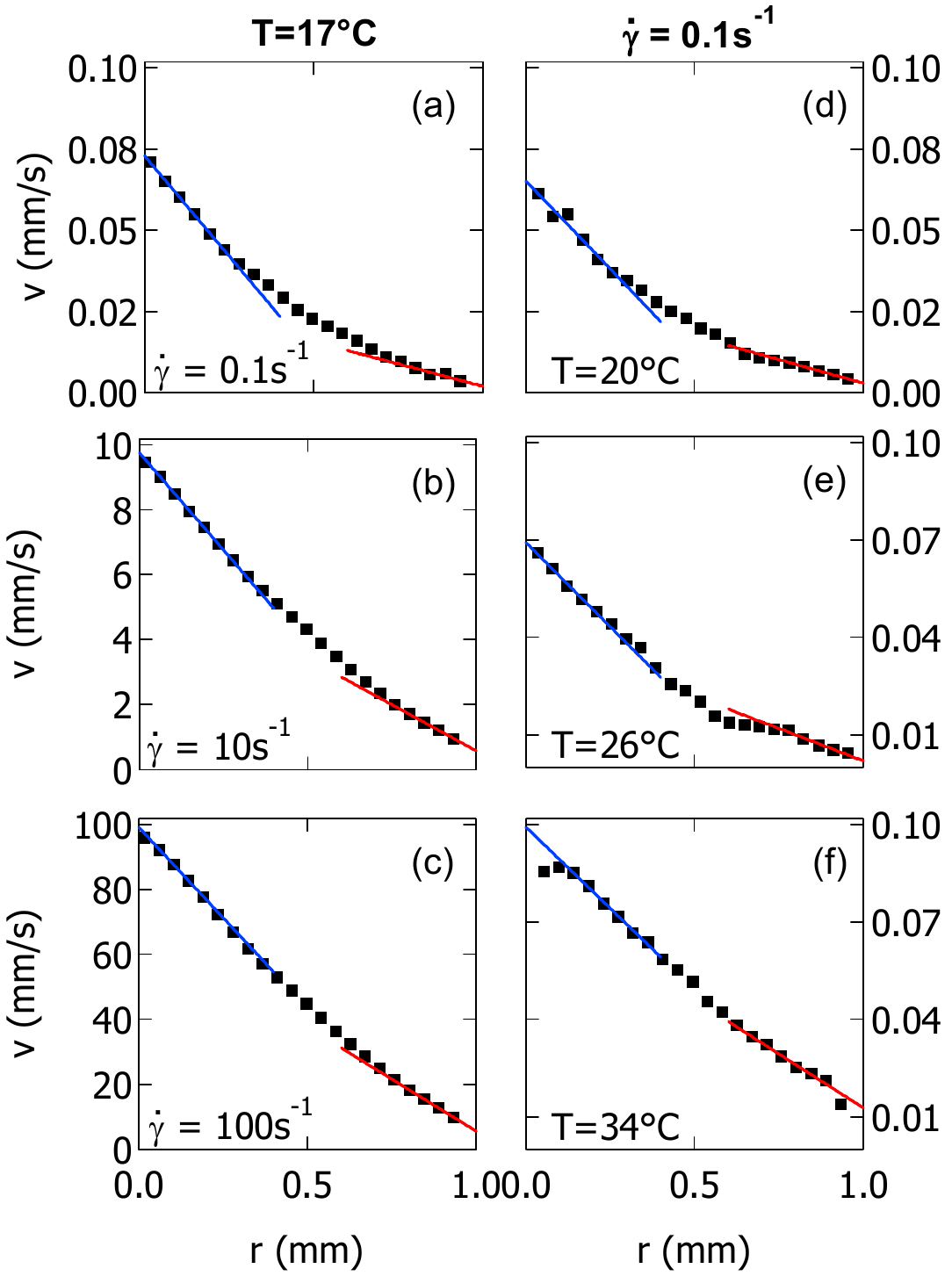}
\caption{(color online) Steady-state velocity profiles $v(r)$ where $r$ is the distance to the moving wall: (a)--(c)~at $T=17^{\circ}$C for different shear rates $\dot \gamma = 0.1$, 10 and 100~s$^{-1}$ from top to bottom and (d)--(f)~at $\dot \gamma = 0.1$~s$^{-1}$ for different temperatures $T=20$, 26 and $34^{\circ}$C from top to bottom. Blue (resp. red) lines are the best linear fits of the velocity profiles over 50 to 150~$\mu$m from the rotor (resp. stator). Extrapolating each fit to the wall gives the fluid velocity $v_f$ at both boundaries. In all cases the upper limit of the vertical axis corresponds to the rotor velocity.
\label{fig2}}
\end{figure} 

\begin{figure*}[!t]
\centering
\includegraphics[width=0.7\linewidth]{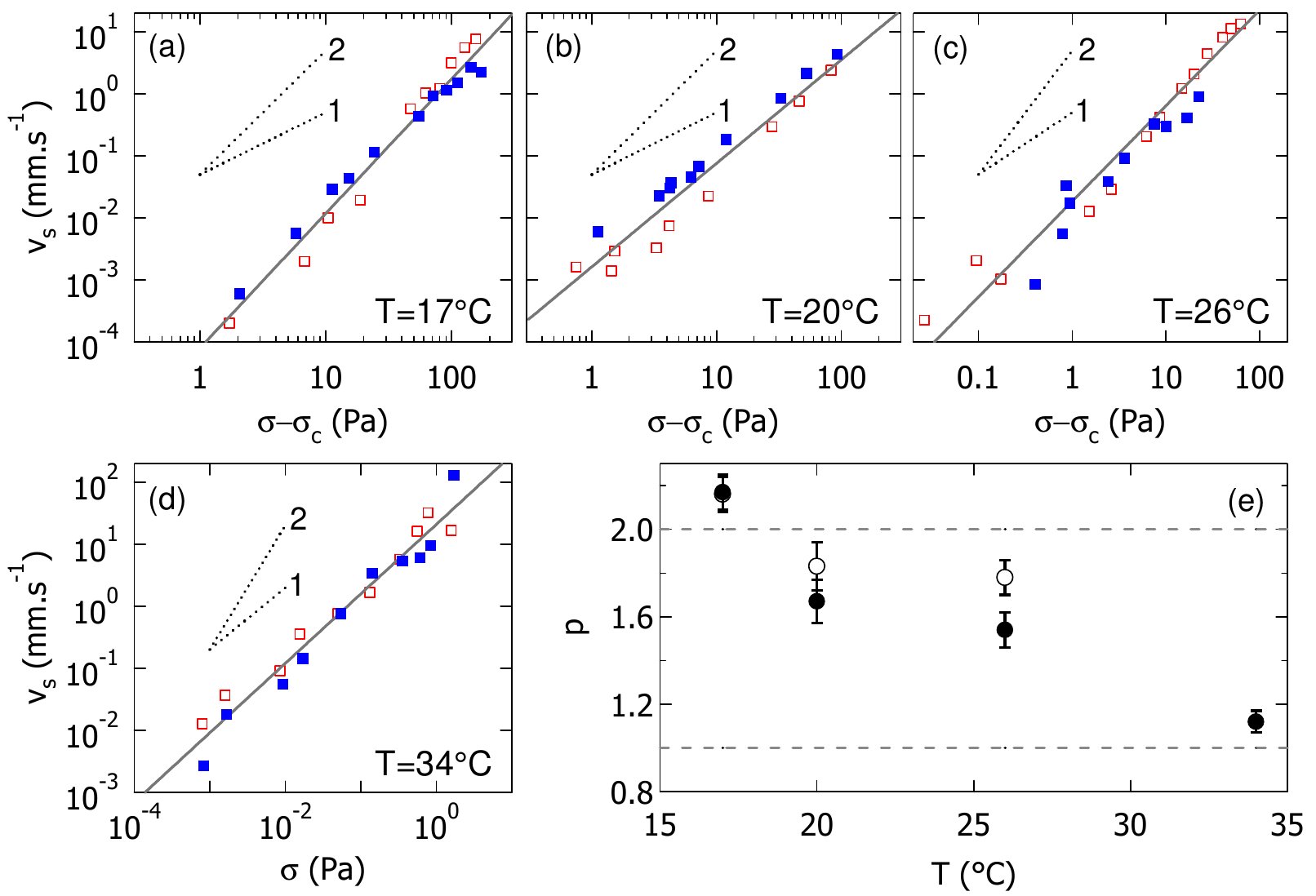}
\caption{(color online) (a)-(d)~Steady-state slip velocities $v_s$ measured at the rotor (\textcolor{red}{$\square$}) and at the stator (\textcolor{blue}{$\blacksquare$}) for different temperatures $T=17$, 20, 26 and 34$^{\circ}$C. Gray lines are the best power-law fits of the data, $v_s=B(\sigma-\sigma_c)^{p}$, where $\sigma_c$ is fixed to the yield stress inferred from the Herschel-Bulkley fit of the flow curve (Fig.~\ref{fig1}) and $B$ and $p$ are free parameters. (e)~Exponent $p$ vs temperature $T$ obtained from the previous fitting procedure ($\bullet$) and by using the value of $\sigma_c$ that minimizes the $\chi^2$ of linear fits of $\ln v_s$ vs $\ln(\sigma-\sigma_c)$ when $\sigma_c$ is varied ($\circ$). In the former case error bars are computed from the uncertainty on $\sigma_c$, whereas in the latter case error bars correspond to the standard deviation associated with the least-square method. The horizontal dashed lines emphasize $p=1$ and 2.
\label{fig3}}
\end{figure*} 

\section{Experimental}

{\it pNIPAM microgels.-} 
Samples consist in poly($N$-isopropylacrylamide) or pNIPAm microgel particles synthesized by aqueous free-radical polymerisation from $N,N'$-methylenebisacrylamide and $N$-isopropylacrylamide (Sigma Aldrich) following a recipe previously described in \cite{Pelton:1986,Wu:1994,Pelton:2000} and more recently applied in \cite{Destribats:2011}. The reaction is performed in a three-neck round-bottom flask (500~mL), equipped with a magnetic stir bar, a reflux condenser, a thermometer and an argon inlet. $N$-isopropylacrylamide (NIPAm) monomers at an initial concentration of 70~mM and $N,N'$-methylenebisacrylamide (BIS) at 2.5~mol\% are dissolved in 98~mL of water. The solution is purified through a 0.2~$\mu$m membrane filter to remove residual particulate matter, before being heated up to 70$^{\circ}$C with argon thoroughly bubbling. Free radical polymerization is initiated after 1~h, with potassium persulfate (KPS, 2.5~mM) dissolved in 2~mL of water. The solution initially transparent becomes progressively turbid due to the polymerization process which is carried out during 6~h in the presence of argon and under stirring. To eliminate chemical residues and water-soluble oligomers, the microgels are then purified by at least 5 centrifugation-redispersion cycles (under 21.$10^3$ $g$ for 1~h, where $g$ denotes the acceleration of gravity). For each cycle, the supernatant is removed and its surface tension is measured by the pendant drop method. The purification process is repeated until the surface tension of the supernatant is comparable to that of pure water, i.e. above 70~mN/m. After the last centrifugation step, the supernatant is totally removed. The microgel concentration is quantified by measuring the dried mass of polymer in the paste and estimated to be equal to 5.2\%~wt. The polydispersity index of the sample was estimated through Dynamic Light Scattering to be constant and about 0.06 in the jammed state ($T\leq 32^{\circ}$C). Such a polydispersity should be enough to ensure that no crystallization is taking place. Finally, the microgel suspensions are seeded with acoustic tracers incorporated at 35$^{\circ}$C, namely 1\%~wt polystyrene microbeads (Dynoseeds TS, 20~$\mu$m diameter, density 1.05), that were checked to have a negligible influence on the system rheology. 

{\it Velocimetry setup.-}
Simple shear experiments are performed in a polished Plexiglas Couette cell (typical roughness 15~nm, height $H=28$~mm, rotating inner cylinder of radius $R=23.96$~mm, fixed outer cylinder of radius 25 mm, gap $e=1.04$~mm) equipped with a homemade lid to minimize evaporation. Temperature is controlled within 0.1$^{\circ}$C by a water circulation around the Couette cell. Rheological data are recorded with a stress-controlled rheometer (MCR 301, Anton Paar) while the azimuthal velocity $v$ is measured simultaneously as a function of the radial distance $r$ to the rotor at about 15~mm from the cell bottom and with a spatial resolution of 40~$\mu$m by means of ultrasonic velocimetry \cite{Manneville:2004}. For each temperature, velocity measurements require the knowledge of $(i)$~the sound speed $c_s$ in the sample, which is deduced from ultrasonic time-of-flight measurements using a standard transmission setup (Table~\ref{table1}) and $(ii)$~the position of the stator--fluid interface and the incidence angle of the ultrasonic beam, that we both determine by calibration in an aqueous dispersion of 1\%~wt polystyrene microbeads \cite{Manneville:2004}.

\section{Results}

{\it Properties of the thermoresponsive colloids.-} The individual size of pNIPAm microgel particles is known to be continuously tunable with  temperature, which can thus be used to vary the packing fraction of the suspension \cite{Zhang:2009,Chen:2011}. Above 34$^{\circ}$C the microgels are collapsed and the sample behaves as a dilute, liquid-like suspension of hard particles with a shear-thinning behaviour [Fig.~\ref{fig1}(a,d)]. As the temperature is decreased the particles swell and their diameter $d$ increases from 300 to 600~nm over a temperature span of about $4^{\circ}$C around the lower critical solution temperature (LCST) $T_{lc}\simeq 32^{\circ}$C [Fig.~\ref{fig1}(a)]. As a result the sample jams into a solidlike material for $T<T_{lc}$ [Fig.~\ref{fig1}(b)] \cite{Carrier:2009,Nordstrom:2010}. The flow curves, shear stress $\sigma$ versus shear rate $\dot \gamma$, show that the jammed suspensions behave as yield stress fluids under shear and are well described by the Herschel-Bulkley model, $\sigma=\sigma_c+A\dot\gamma^n$, with a yield stress $\sigma_c$ that steeply decreases as $T$ approaches $T_{lc}$ [Fig.~\ref{fig1}(c) and Table~\ref{table1}]. 

{\it Slip velocity measurements.-} Once the sample is loaded into the shear cell, we wait for its viscoelastic moduli to reach a plateau. We then impose various shear rates $\dot \gamma$ and record the velocity profiles together with the stress response before moving to another temperature and repeating the same protocol. 
At a given temperature, two consecutive shear start-up experiments are separated by a preshear at 50~s$^{-1}$ during 40~s to erase the sample history, followed by a 3~min rest period during which we monitor the evolution of $G'$ and $G''$ through small amplitude oscillatory shear. Quantitative agreement in the viscoelastic moduli measured between successive shear experiments indicates that the sample neither ages nor suffers from evaporation issues.
Steady-state velocity profiles obtained at $T=17^{\circ}$C for different shear rates are reported in Fig.~\ref{fig2}(a--c). The choice of smooth boundary conditions favors wall slip over a large range of shear stresses as evidenced by the difference between the extrapolation of the velocity profile at the rotor and the rotor velocity imposed by the rheometer. Note that the bulk sample is sheared in all the cases reported here, proving that we only consider stresses larger than the yield stress in contrast with previous work \cite{Meeker:2004a,Meeker:2004b}. Linear fits of the velocity profiles over 100~$\mu$m at a distance of 50~$\mu$m from the walls allow us to estimate the fluid velocity at the rotor and the stator, respectively noted $v_{f}^{(r)}$ and $v_{f}^{(s)}$. Slip velocities are then computed as $v_s^{(i)}=\vert v_{f}^{(i)}-v_{w}^{(i)}\vert$, with $i=r$ or $s$, and where $v_{w}^{(r)} =\Omega R$ and $v_{w}^{(s)}=0$, with $\Omega$ the rotational speed of the rotor.

\begin{table}[t!]
\begin{tabular}{c|c|c|c|c|c}
$T$ ($^{\circ}$C) & $c_s$ (m.s$^{-1}$) & $\sigma_c$ (Pa) & $n$ & $A$ (Pa.s$^n$) & $\sigma_c^{(\chi^2)}$ (Pa) \\
\hline\hline
17 & 1533$\pm$4 & 45.7$\pm$0.2 & 0.44$\pm$0.02 & 15.5 $\pm$0.2 & 45.8$\pm$0.1\\ 
20 & 1536$\pm$4 & 31.3$\pm$0.4 & 0.49$\pm$0.02 & 8.4 $\pm$0.2 & 30.8$\pm$0.1\\
26 & 1541$\pm$5 & 8.9$\pm$0.1 & 0.54$\pm$0.02 & 2.33$\pm$0.07 & 8.7$\pm$0.05\\
34 & 1548$\pm$5 & 0 		  & 0.74$\pm$0.02 & 0.23$\pm$0.01 & 0\\
\end{tabular}
\caption{Sound speed and rheological parameters of the pNIPAm microgel suspensions at different temperatures $T$. The last column shows the value of $\sigma_c$ inferred from the least-square procedure detailed in the main text.} \label{table1}
\end{table}

The dependence of $v_s^{(r)}$ and $v_s^{(s)}$ on the stress at the wall is reported in Fig.~\ref{fig3}(a). Slip velocities are plotted as a function of the viscous stress, $\sigma -\sigma_c$, where $\sigma_c$ is fixed to the value of the yield stress inferred from the Herschel-Bulkley fit of the steady-state flow curve shown in Fig.~\ref{fig1}(c). Moreover, the local stress in the Taylor-Couette geometry is given by $\sigma(r)=\Gamma/[2\pi H(R+r)^2]$, where $\Gamma$ is the torque on the rotor, so that in our geometry the stress decreases by about $2e/R\simeq 8$\% from the rotor to the stator. Such a variation is taken into account in Fig.~\ref{fig3} by plotting $v_s^{(r)}$ and $v_s^{(s)}$ as a function of $\sigma^{(r)}=\sigma(0)$ and $\sigma^{(s)}=\sigma(e)$ respectively.
 As a first key observation both slip velocities increase as a power law of the viscous stress over two decades of stress and four decades of velocities, with an exponent $p=2.17\pm 0.08$. This result strikingly contrasts with previous analyses in which $v_s$ was reported as a function of $\sigma$ rather than $\sigma -\sigma_c$ \cite{Salmon:2003,Geraud:2013,Seth:2012,PerezGonzalez:2012,Poumaere:2014}. We have repeated these measurements at different temperatures above, around and below the jamming transition, respectively at $T=20$, $26$ and $34^{\circ}$C [Fig~\ref{fig3}(b--d)]. For all temperatures $v_s$ robustly scales as a power law of the viscous stress (taking $\sigma_c=0$ for the liquidlike suspension at $T=34^{\circ}$C). Furthermore, the exponent $p$ is found to be the same at the rotor and at the stator within error bars, and to decrease smoothly from 2 to 1 for increasing temperatures, taking the intermediate value $p = 1.54 \pm 0.08$ for $T=26^{\circ}$C [Fig.~\ref{fig3}(e)]. The determination of $p$ being sensitive to the value of $\sigma_c$, we have performed the following test to confirm the relevance of the above power-law scaling: instead of using the yield stress extracted from the Herschel-Bulkley fit of the flow curve, $\sigma_c$ is chosen as the value that minimizes the $\chi ^2$ of linear fits of $\ln v_s$ vs $\ln(\sigma-\sigma_c)$ when $\sigma_c$ is varied. The corresponding estimates are in good agreement with the previous values of $p$ [Fig.~\ref{fig3}(e)]. Furthermore the values of $\sigma_c$ inferred from the least-square method are compatible within error bars with those from the Herschel-Bulkley fit (Table~\ref{table1}).

\begin{figure}[!t]
\centering
\includegraphics[width=\linewidth]{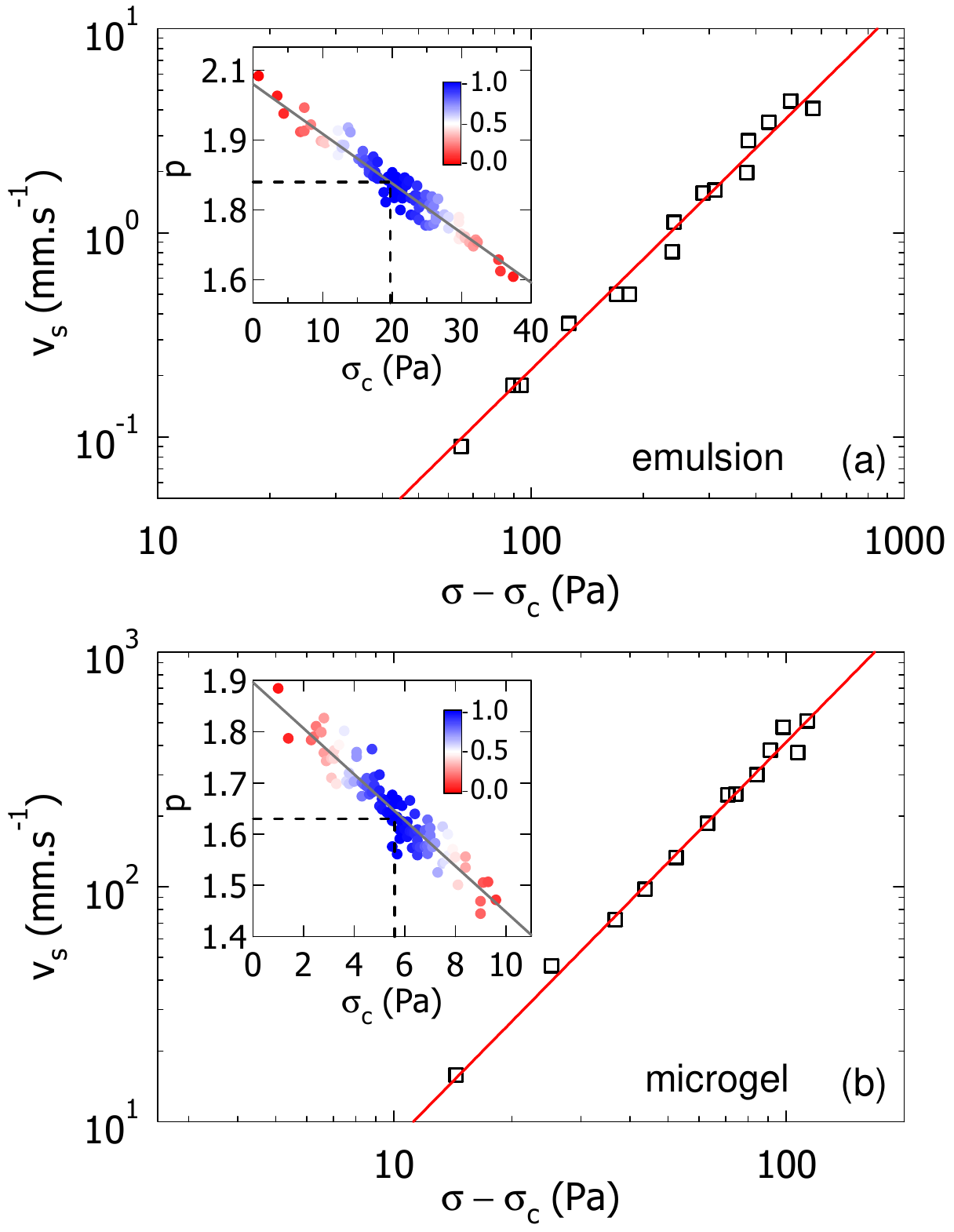}
\caption{(color online) Steady state slip velocity $v_s$ vs the viscous stress $\sigma-\sigma_c$ for (a)~a concentrated emulsion ($\phi \simeq 0.75$) sheared in a 3~mm gap Taylor Couette cell (data from \cite{Salmon:2003}) and (b)~a 1\%~wt. carbopol microgel sheared in a rough microchannel (data from \cite{Geraud:2013}). In the absence of independent reliable estimates for $\sigma_c$, we used the yield stress obtained by minimizing the $\chi ^2$ of linear fits of $\ln v_s$ vs $\ln(\sigma-\sigma_c)$, respectively $\sigma_c=19.7$~Pa in the emulsion and $\sigma_c=5.5$~Pa in the microgel. The red lines are the best power-law fits leading to $p=1.81 \pm 0.06$ and $p=1.63 \pm 0.05$ respectively. Insets in (a) and (b) show the exponents $p$ vs the yield stress $\sigma_c$ inferred from a hundred least-square procedures on the same data sets to which a uniform noise of 10~\% amplitude has been added. The dashed lines indicate the pair $(\sigma_c,p)$ obtained in the absence of noise. Colors code for the probability distribution of the yield stress values. 
\label{fig.4}}
\end{figure} 

\section{Discussion and conclusion}
We have shown that the power-law scaling $v_s \propto (\sigma-\sigma_c)^p$ robustly describes the slip behavior of jammed suspensions of thermoresponsive pNIPAm particles sheared above $\sigma_c$. This prompts us to revisit two data sets where $v_s$ was originally reported to scale as $\sigma^2$: ($i$)~a concentrated emulsion sheared in a Taylor-Couette cell with smooth boundary conditions \cite{Salmon:2003} and ($ii$)~a carbopol microgel sheared in a microchannel with rough boundary conditions \cite{Geraud:2013}. Figure~\ref{fig.4} shows that both data sets are very well described by power laws of the viscous stress with $p=1.81 \pm 0.06$ (emulsion) and $p=1.63 \pm 0.05$ (microgel). To further discriminate between $v_s$ being a function of the stress or of the viscous stress, we determine the pairs of best fit parameters $(\sigma_c,p)$ from a hundred realizations of the $\chi^2$ minimization procedure where a uniform noise of 10\% amplitude has been added to the data (insets of Fig.~\ref{fig.4}). In both cases the pair $(\sigma_c=0,p\simeq 2)$ lies at the very edge of the distribution, which allows us to rule out a quadratic scaling with $\sigma$. Our work thus provides an explanation to reconcile previous inconsistent results for stresses larger than the yield stress. It also draws a link with the slip behaviour of soft jammed materials {\it below} $\sigma_c$ where the material is solidlike and flows as a plug. Indeed, in microgels and emulsions for $\sigma<\sigma_c$, $v_s$ was shown to increase either linearly or quadratically, i.e. $v_s \propto (\sigma-\sigma_s)^p$, where $\sigma_s$ is the stress below which the material sticks to the boundaries, with $p=1$ or 2 depending on the nature of the interactions with the wall \cite{Meeker:2004a,Meeker:2004b,Seth:2008,Seth:2012}. Our results show that for $\sigma > \sigma_c$ the apparent yield stress $\sigma_s$ is replaced by the bulk yield stress $\sigma_c$ and that the exponent $p$ depends on the packing fraction. 
The latter result can be understood as follows: at low temperatures compared to the LCST, the exponent $p=2$ can be interpreted as the result of a force balance on compressed particles experiencing Hertz-like deformation and lubrication forces with the walls, as first proposed for plug flows in \cite{Meeker:2004a,Meeker:2004b}. At intermediate temperatures closer to the LCST, such a framework is probably no longer valid as the microgels may experience not only lubrication forces but also cooperative rearrangements thanks to both the lower volume fraction and the non-zero local shear rate. Let us indeed recall that the material is sheared in the bulk for all data reported here, whatever the shear rate or the temperature.  The lesser importance of the lubrication interactions for decreasing packing fractions could thus account for the decrease of the exponent $p$, and for the transition to dilute-like behavior of the suspension, although the packing fraction is still above jamming.

To conclude, we have shown that slip velocities in yield stress fluids made of soft jammed particles and sheared in the bulk increase as power laws of the viscous stress over a range of temperatures that encompasses both sides of the jamming transition. Therefore $v_s$ can be written under the single form $v_s \propto [\sigma-\sigma_c(\phi)]^{p(\phi)}$ with $\sigma_c=0$ and $p\simeq 1$ below jamming and where both $\sigma_c$ and $p$ are increasing functions of $\phi$ above jamming. These results hint at a link between the bulk behavior of the material and the slip at the wall, as also evidenced by some of the nonlinear velocity profiles reported in Fig.~\ref{fig2}. Whether such nonlinearity should be attributed to some flow-concentration coupling \cite{Besseling:2010} or to long-range effects of the boundaries associated with non-local effects \cite{Seth:2012,Bocquet:2009,Mansard:2012} remains an open issue. In any case a complete description of the present data shall certainly involve some coupling between bulk and boundary effects. Such a point of view contrasts with the interpretation of wall slip in foams, where slippage is mainly controlled by the physics at the boundaries and decoupled from bulk properties \cite{Saugey:2006,Cantat:2013,LeMerrer:2015}. Clearly a more detailed study of boundary-driven vs bulk-driven effects involving wall slip remains to be conducted by collecting data on various soft glassy materials including foams. Finally, our work paves the way for a more realistic modeling of flows of dense suspensions in the presence of slip. Spatially resolved model such as SGR and fluidity models \cite{Fielding:2014} should focus on the scaling of $v_s$ with $\sigma$ before turning to transient regimes for which wall slip goes hand in hand with long-lived heterogeneous dynamics \cite{Divoux:2012,Perge:2014}.\\  

\begin{acknowledgments}
The authors thank V.~Grenard for his help with preliminary experiments and C.~Barentin, I.~Cantat, M.~Cloitre, S.~Cohen-Addad, J.~Goyon and G.~Ovarlez for fruitful discussions. This work was funded by the Institut Universitaire de France, the IdEx Bordeaux and the CNRS through the ``Initial Support for Exploratory Projects" (PEPS) scheme (ANR-10-IDEX-03-02), and the European Research Council under the European Union's Seventh Framework Programme (FP7/2007-2013)/ERC Grant Agreement No. 258803.
\end{acknowledgments}

\end{document}